\documentclass[a4paper,11pt]{article}
\usepackage{jhepmod}
\usepackage{lineno}
\usepackage{amsmath,amssymb,mathtools}
\usepackage{xcolor}
\usepackage{array,caption,needspace}
\numberwithin{equation}{section}

\title{Ultraviolet Structure of the Gravitational-Wave Response in a Time-Dependent Scalar Medium}

\author{Han Lai and Atsuhisa Ota}
\affiliation{Department of Physics and Chongqing Key Laboratory for Strongly Coupled Physics, \\
Chongqing University, Chongqing 401331, People's Republic of China}

\emailAdd{aota@cqu.edu.cn}

\abstract{
We study the real-time linear response of gravitational waves in an FLRW background with a scalar field of general time-dependent effective mass.
In the Schwinger-Keldysh formalism, we develop an adiabatic regularization scheme for unequal-time correlation functions and use it to extract the ultraviolet structure of the one-loop response.
The leading divergence reproduces the familiar $\Box ^2h_{ij}$ structure, whereas the time-dependent background induces additional local divergences proportional to $\Box h_{ij}$, $\partial_\tau h_{ij}$, and $h_{ij}$.
These are renormalized by local counterterms associated with the Weyl-squared term, a time-dependent Ricci-scalar term, a time-dependent cosmological constant, and the Ricci-scalar-squared term $R^2$.
The same counterterms also remove the ultraviolet divergences of the background tadpole stress tensor and agree with the known covariant counterterm results.
In a flat-background Mathieu oscillator example, replacing the adiabatic vacuum by the instantaneous vacuum leaves these counterterms unchanged but introduces an initial-surface divergence, which is absorbed into the initial data rather than the bulk counterterms.
}

\begin{document}
\maketitle
\flushbottom

\section{Introduction}\label{frw1:intro}

Parametric resonance after cosmic inflation provides a natural mechanism for transferring energy from the inflaton sector to other degrees of freedom, thereby initiating the hot Big Bang phase~\cite{Kofman:1994rk,Kofman:1997yn,Shtanov:1994ce,Kaiser:1995fb}. During this stage the background is far from equilibrium, strongly time dependent, and parametrically unstable. Particle production can be efficient, and the resulting state need not admit a simple quasiparticle description. In this paper we are interested in the propagation of gravitational waves in such a background and, more generally, in a scalar medium with a time-dependent mass.

This problem should be distinguished from the production of gravitational waves by the preheating source itself~\cite{Khlebnikov:1997di,Easther:2006vd,Garcia-Bellido:2007nns,Dufaux:2007pt}. Our concern is the real-time linear response of gravitational waves to a scalar field with a time-dependent effective mass in an FLRW background. In this respect the problem is closer in spirit to the propagation of gravitational waves in a medium, as in the familiar example of a radiation-dominated universe with free-streaming neutrinos~\cite{Rebhan:1994zw,Weinberg:2003ur,Ota:2023iyh,Ota:2025rll,Ota:2026yzp}. In the resonant case, particle production is ongoing, the state is nonstationary, and a real-time analysis is unavoidable. We would like to understand whether gravitational waves propagating through a resonantly evolving field exhibit features absent in a medium of freely propagating particles. A scalar field with a time-dependent mass provides a minimal setting for this comparison.

The first question that must be settled is the ultraviolet structure of the theory. Because the initial state is the vacuum, the ultraviolet behavior is not softened by thermal occupation numbers as in the case of the neutrino example. Whatever nonlocal effects may appear in the real-time response, they can be discussed only after the divergent local part has been identified and removed. The main purpose of this paper is to carry out that analysis.

We formulate the problem in the Schwinger-Keldysh formalism~\cite{Calzetta:1986cq,Calzetta:1986ey,Kamenev:2009jj} and renormalize the one-loop gravitational-wave response in a time-dependent background. In highly symmetric backgrounds such as Minkowski or de Sitter space, the ultraviolet structure is more tightly constrained and can be organized in terms of standard local counterterms with constant coefficients~\cite{Anderson:2002fk,Marunovic:2011zw,Weinberg:2005vy,Senatore:2009cf}. In contrast, the breaking of time-translation invariance in a time-dependent background leads to a more intricate, time-dependent counterterm structure. For the time-dependent FLRW background considered here, the covariant local counterterm structures are known from the DeWitt--Schwinger expansion~\cite{Vassilevich:2003xt}. This expansion, however, does not by itself separate the divergences from the finite, nonlocal part of the real-time response in an initial-value problem. We perform this separation within the response kernel, by an asymptotic expansion of the unequal-time correlators in the spirit of adiabatic regularization, and the agreement with the known counterterms then serves as a check.

To extract the ultraviolet structure, we combine the adiabatic expansion~\cite{Parker:1974qw,Birrell:1982ix} with a large-momentum expansion at fixed momentum times time separation, which we call the adiabatic-ultraviolet expansion. At leading ultraviolet order, our results reproduce the familiar ultraviolet structure known from semiclassical gravity and linear-response analyses around Minkowski space~\cite{Anderson:2002fk,Marunovic:2011zw}. Earlier work on renormalization in preheating backgrounds~\cite{Armendariz-Picon:2019csc,Armendariz-Picon:2020tkc} focused, respectively, on the energy density of gravitational waves produced by the source and on the background backreaction, while ref.~\cite{Bassett:1997ke} emphasized a Mathieu-type structure in a particular treatment of the Weyl tensor. Our concern here is the local ultraviolet structure of the gravitational-wave linear response in a time-dependent scalar medium on an FLRW background.

We consider a scalar field with a time-dependent effective mass in FLRW spacetime, where the effective mass squared $U(\tau)$ is a general smooth, spatially homogeneous function. In this setting the ultraviolet structure can be analyzed explicitly, and the required counterterms can be identified in closed form. We then compare the renormalization of the linear response with that of the tadpole stress tensor. Once the contact response is included, the same counterterms remove the local ultraviolet divergences in both. In a flat-background Mathieu oscillator example, we also compare the adiabatic vacuum with the instantaneous vacuum: the bulk counterterms are unchanged, but the latter introduces an initial-surface divergence, which is absorbed into the initial data.

This paper is organized as follows. In section~\ref{frw23:setup} we set up the problem and derive the real-time linear response of gravitational waves to a scalar field with a time-dependent effective mass in an FLRW background. In section~\ref{frw23:adiabatic} we develop the adiabatic-ultraviolet expansion appropriate to the unequal-time Green functions that enter the response kernel. In section~\ref{frw4:section} we determine the ultraviolet structure of the one-loop response and identify the local counterterms required for renormalization. In section~\ref{frw5:sec} we examine the relation between the renormalization of the linear response and that of the tadpole stress tensor. In section~\ref{frw6:resonance} we apply the general results to a parametric resonance background, giving the explicit divergent structure, comparing the adiabatic and instantaneous vacua, and illustrating Pauli--Villars regularization of the tadpole. In section~\ref{frw6:concl} we summarize our conclusions.

\section{Setup}\label{frw23:setup}

In this section we formulate the real-time linear-response problem for gravitational waves in an FLRW background with a scalar field of time-dependent mass.
\subsection{Quantization of a scalar field with time-dependent mass}

Let us consider a scalar field $\chi$ with a time-dependent effective mass squared $U(\tau)$:
\begin{align}
S_\chi=-\frac12\int d^4x\sqrt{-g}
\left[g^{\mu\nu}\partial_\mu\chi\partial_\nu\chi+U(\tau)\chi^2\right].
\label{frw23:scalar-action}
\end{align}
Here $U$ is a general smooth, spatially homogeneous effective mass squared and $\tau$ is conformal time.

The effective mass may arise from the coupling of the scalar field to a spatially homogeneous background. For example, during preheating, if the interaction term between $\chi$ and a background scalar field $\phi$ is $-g^2\phi^2\chi^2/2$, then $U=g^2\phi^2$, where $g$ is the coupling constant. Taking a coherently oscillating background $\phi(t)=\Phi(t)\sin(mt)$ gives $U(\tau)=g^2\Phi^2(t(\tau))\sin^2[mt(\tau)]$; here $t$ is cosmic time related to conformal time by $dt=a\,d\tau$, $\Phi$ is the amplitude, and $m$ is the mass of the $\phi$ field and also its coherent oscillation frequency. In what follows we do not restrict ourselves to this particular form.
We study the local ultraviolet structure arising as the loop momentum $p\to\infty$ for a fixed smooth background and fixed external momentum $k$.

Let us consider a spatially flat FLRW background 
\begin{align}
g^{\mu\nu}=a^{-2}(\tau)\eta^{\mu\nu},\quad
\mathcal H=\frac{a'}a,\quad R_0=\frac{6a''}{a^3}.
\label{frw23:frw-background}
\end{align}
Here a prime denotes a derivative with respect to conformal time, and $R_0$ is the background Ricci scalar. We use the mostly-plus signature, $\eta^{\mu\nu}=\mathrm{diag}(-1,1,1,1)$.
For convenience, we rescale the field as $\psi=a\chi$.
The action then takes the form
\begin{align}
S_\chi^{(0)}
&=\int d\tau\,d^3x
\mathcal L^{(0)}_{\psi}
,\label{frw23:background-rescaled-action}
\\
\mathcal L^{(0)}_{\psi}&=\frac12\left[
(\psi'-\mathcal H\psi)^2-(\partial_i\psi)^2-a^2U\psi^2\right].
\label{frw23:raw-lagrangian}
\end{align}
Denoting the conjugate momentum of $\psi$ by $\pi_\psi\equiv \psi'-\mathcal H\psi$, we impose the equal-time commutation relation
\begin{align}
[\psi(\tau,\vec x),\pi_\psi(\tau,\vec y)]
=i\delta^{(3)}(\vec x-\vec y).
\end{align}
In a time-dependent background, one may Fourier transform only in space:
\begin{align}
\psi(\tau,\vec x)
&=\int\frac{d^3p}{(2\pi)^3}e^{i\vec p\cdot\vec x}
\left[a_{\vec p}(\tau_0)u_p(\tau,\tau_0)
+a^\dagger_{-\vec p}(\tau_0)u_p^*(\tau,\tau_0)\right],
\\
\pi_\psi(\tau,\vec x)
&=\int\frac{d^3p}{(2\pi)^3}e^{i\vec p\cdot\vec x}
\left[a_{\vec p}(\tau_0)v_p(\tau,\tau_0)
+a^\dagger_{-\vec p}(\tau_0)v_p^*(\tau,\tau_0)\right],
\end{align}
where the mode function for the conjugate momentum is defined as
\begin{align}
    v_p \equiv u_p'-\mathcal H u_p
\end{align}
The annihilation and creation operators satisfy
\begin{align}
[a_{\vec p},a^\dagger_{\vec q}]=(2\pi)^3\delta^{(3)}(\vec p-\vec q),
\end{align}
and the mode functions satisfy the Wronskian condition
\begin{align}
    u_pv_p^*-v_pu_p^*=i.
\end{align}
Because of the explicit time dependence of the background, the Hamiltonian is not conserved, and the definition of the vacuum also depends on the choice of initial time.
In this paper, we define $|0\rangle$ such that $a_{\vec p}(\tau_0)|0\rangle=0$, with the mode functions specified by the adiabatic expansion.
This adiabatic vacuum differs in general from the instantaneous vacuum defined by diagonalizing the Hamiltonian at the initial time.
We will discuss the physical implications of this choice in section~\ref{frw6:vacua}.

The mode function $u_p$ satisfies
\begin{align}
u_p''+\left(\omega_p^2-\frac{a''}a\right)u_p=0,\qquad
\omega_p^2(\tau)\equiv p^2+a^2(\tau)U(\tau).
\label{frw23:mode-equation}
\end{align}
In what follows, $u_p$ always denotes the mode function evolving according to Eq.~\eqref{frw23:mode-equation}. The ultraviolet conditions on the initial state are given in section~\ref{frw23:adiabatic}.

In interaction-picture perturbation theory in the Schwinger--Keldysh formalism, the basic building blocks of the response functions are the retarded and Keldysh Green's functions~\cite{Chou:1984es,Berges:2004yj,Calzetta:2008iqa}:
\begin{align}
G_p^R(\tau,\tau')
&=i\Theta(\tau-\tau')
\left[u_p(\tau)u_p^*(\tau')-u_p^*(\tau)u_p(\tau')\right],
\label{frw23:gr}\\
G_p^K(\tau,\tau')
&=u_p(\tau)u_p^*(\tau')+u_p(\tau')u_p^*(\tau).
\label{frw23:gk}
\end{align}
These are the Green functions of the field variable $\psi$; the two-point functions of the original field $\chi$ contain an additional factor $[a(\tau)a(\tau')]^{-1}$.

\subsection{Gravitational waves and their linear response}

Gravitational waves are defined as transverse-traceless~(TT) metric perturbations:
\begin{align}
g^{\mu\nu}=a^{-2}(\eta^{\mu\nu}+\kappa h^{\mu\nu}),\qquad
h^{0\mu}=0,\qquad \delta_{ij}h^{ij}=\partial_i h^{ij}=0,
\label{frw23:metric-convention}
\end{align}
Here $\kappa=\sqrt{32\pi G}$, with $G$ Newton's constant. Spatial indices of the perturbation are raised and lowered with $\delta_{ij}$, and the d'Alembertian associated with the flat metric $\eta_{\mu\nu}$ is denoted by $\Box=-\eta^{\mu\nu}\partial_\mu\partial_\nu=\partial_\tau^2-\nabla^2$.
The metric determinant expands as
\begin{align}
\sqrt{-g}=a^4\left(1+\frac{\kappa^2}{4}h^{ij}h_{ij}\right)+O(h^3).
\label{frw23:determinant}
\end{align}
Substituting Eqs.~\eqref{frw23:metric-convention} and \eqref{frw23:determinant} into Eq.~\eqref{frw23:scalar-action}, we find
\begin{align}
S_\chi&=\int d^4x\left[
\mathcal L^{(0)}_{\psi}-\frac\kappa2h^{ij}\partial_i\psi\partial_j\psi
+\frac{\kappa^2}{4}h^{ij}h_{ij}\mathcal L^{(0)}_{\psi}\right]+O(h^3).
\label{frw23:rescaled-full-action}
\end{align}
The linear coupling arises from the kinetic term of the scalar field and couples the first-order metric perturbation to the unperturbed stress tensor. The traceless condition on $h^{ij}$ projects this coupling onto $\partial_i\psi\partial_j\psi$. The interaction Hamiltonian is then
\begin{align}
H_{\rm int}(\tau)=\frac\kappa2\int d^3x\,
h^{ij}\partial_i\psi\partial_j\psi.
\end{align}
The dynamical response is, up to the TT projection,
\begin{align}
a^2T^{\rm dyn}_{ij}(\tau,\vec x)
=i\int_{\tau_0}^{\tau}d\tau'\,
\langle0|[H_{\rm int}(\tau'),
\partial_i\psi(\tau,\vec x)\partial_j\psi(\tau,\vec x)]|0\rangle.
\end{align}
A straightforward algebraic manipulation gives~\cite{Ota:2025rll}
\begin{align}
a^2T^{\rm dyn}_{ij}(\tau,\vec k)
={}&\kappa\int_{\tau_0}^{\tau}d\tau'\int_0^\infty dp
\int_{|p-k|}^{p+k}dp'\,w(k,p,p') G^R_{p'}(\tau,\tau')G^K_p(\tau,\tau')
h_{ij}(\tau',\vec k),
\label{frw23:response-kernel}\\
w(k,p,p')={}&-\frac{pp'\left[k^4-2k^2(p^2+p'^2)+(p^2-p'^2)^2\right]^2}
{256\pi^2k^5}.
\label{frw23:angular-weight}
\end{align}
In addition to the dynamical response above, the explicit metric dependence of the stress tensor
\begin{align}
a^2T_{ij}
=&\partial_i\psi\,\partial_j\psi
+\delta_{ij}\mathcal L^{(0)}_\psi
-\kappa h_{ij}\mathcal L^{(0)}_\psi
-\frac{\kappa}{2}\delta_{ij}
h^{kl}\partial_k\psi\,\partial_l\psi
+O(\kappa^2).
\end{align}
gives rise to a contact response:
\begin{align}
T^{\rm cont}_{ij}(x)\equiv \int d^4 y\,
\left\langle 0\left|\left (
\frac{\delta T_{ij}(x)}{\kappa\delta h^{k\ell}(y)}
\right) _{\kappa=0}\right|0\right\rangle \kappa h^{k\ell}(y).
\end{align}
We then obtain the contact contribution
\begin{align}
a^2T^{\rm cont,TT}_{ij}=-\kappa h_{ij}\langle 0 |\mathcal L^{(0)}_{\psi}|0 \rangle.
\label{frw23:contact-operator}
\end{align}
In the unperturbed background, $\langle 0 |\partial_i\psi\,\partial_j\psi|0 \rangle \propto \delta_{ij}$ and therefore vanishes under TT projection. Its first-order dynamical response need not vanish.
The full response is therefore
\begin{align}
a^2T^{\rm lin, TT}_{ij}
=a^2\left[T^{\rm dyn}_{ij}+T^{\rm cont}_{ij}\right]^{\rm TT},
\label{frw23:full-response}
\end{align}
which enters the right-hand side of the linearized Einstein equation and describes the linear response of gravitational waves in the scalar field background. Since $G^K$ contains vacuum fluctuations, this response is ultraviolet divergent.

\section{Adiabatic-ultraviolet expansion}\label{frw23:adiabatic}

To extract the ultraviolet structure of the unequal-time response kernel~\eqref{frw23:response-kernel}, we construct an adiabatic-ultraviolet expansion of the mode functions. Although analytic closed-form expressions are not available in a general time-dependent background, the adiabatic asymptotic expansion of high-momentum modes captures the short-distance structure relevant for renormalization~\cite{Parker:1974qw,Birrell:1982ix}. We begin with the adiabatic approximation and then combine it with the short-time expansion required for unequal-time correlators.

\subsection{Adiabatic approximation}
\label{adiabaticsection}

We introduce the zeroth-order frequency
\begin{align}
\omega_p^2=p^2+a^2U.
\label{frw23:omega0}
\end{align}
The curvature term $-a''/a$ contains two time derivatives and is counted as second order, and we assume that $U$ does not contain time derivatives. We denote the adiabatic approximation to the mode function by $u_p^{\rm A}$. Suppose that the time dependence of $\omega_p(\tau)$ is negligible. At zeroth adiabatic order, the mode function is given by
\begin{align}
u_p^{\rm A}(\tau,\tau_0)
\sim\frac1{\sqrt{2\omega_p}}e^{-i\omega_p(\tau-\tau_0)}.
\label{frw23:constant-frequency-mode}
\end{align}
In the adiabatic approximation, we generalize this ansatz to
\begin{align}
u_p^{\rm A}(\tau,\tau_0)
=\frac1{\sqrt{2W_p(\tau)}}
\exp\left[-i\int_{\tau_0}^{\tau}W_p(s)\,ds\right].
\label{frw23:wkb-ansatz}
\end{align}
We determine $W_p$ self-consistently in an expansion in time derivatives. Substituting this ansatz into Eq.~\eqref{frw23:mode-equation} and using $\lambda$ to count time derivatives, we find
\begin{align}
W_p^2=\omega_p^2-\lambda^2\frac{a''}a
+\lambda^2\frac34\frac{W_p'^2}{W_p^2}
-\lambda^2\frac{W_p''}{2W_p}.
\label{frw23:riccati}
\end{align}
Here $\lambda$ is a bookkeeping parameter, and the physical result is obtained by setting $\lambda=1$. Writing
\begin{align}
W_p=W_{0,p}+\lambda^2W_{2,p}+\lambda^4W_{4,p}+\cdots,
\qquad W_{0,p}=\omega_p,
\end{align}
we substitute this expansion into Eq.~\eqref{frw23:riccati} and solve order by order in $\lambda$. The left-hand side expands as
\begin{align}
W_p^2=W_{0,p}^2+2\lambda^2W_{0,p}W_{2,p}
+\lambda^4\left(2W_{0,p}W_{4,p}+W_{2,p}^2\right)
+O(\lambda^6).
\end{align}
We thus obtain
\begin{align}
W_{2,p}
&=-\frac{a''}{2a\omega_p}
+\frac{3\omega_p'^2-2\omega_p\omega_p''}{8\omega_p^3},
\label{frw23:w2}\\
W_{4,p}
&=-\frac{W_{2,p}^2}{2\omega_p}
-\frac{W_{2,p}''}{4\omega_p^2}
+\frac{3\omega_p'W_{2,p}'+\omega_p''W_{2,p}}{4\omega_p^3}
-\frac{3\omega_p'^2W_{2,p}}{4\omega_p^4}.
\label{frw23:w4}
\end{align}
In this way one obtains the adiabatic approximation to the mode function. The adiabatic expansion counts time derivatives, and its applicability requires the following quantities to be sufficiently small:
\begin{align}
\left|\frac{a''}{a\omega_p^2}\right|\ll1,\qquad
\frac{\omega_p'^2}{\omega_p^4}\ll1,\qquad
\left|\frac{\omega_p''}{\omega_p^3}\right|\ll1.
\end{align}
From Eqs.~\eqref{frw23:w2} and~\eqref{frw23:w4}, we have
\begin{align}
W_{2,p}&=-\frac{a''}{2ap}+O(p^{-3}),\notag\\
W_{4,p}&=\frac{(a''/a)''-(a''/a)^2}{8p^3}+O(p^{-5}).
\end{align}
The recursion relation gives $W_{2j,p}=O(p^{1-2j})$ ($j\geq1$), and their asymptotic behavior is
\begin{align}
W_{2,p}=O(p^{-1}),\qquad W_{4,p}=O(p^{-3}),\qquad
W_{6,p}=O(p^{-5}).
\label{frw23:W-order}
\end{align}
In what follows, we use
\begin{align}
W_p^{[4]}=W_{0,p}+W_{2,p}+W_{4,p}.
\end{align}
Its sufficiency will be explained below by considering both the amplitude and the short-time phase.

\subsection{Ultraviolet expansion}

We use the adiabatic expansion to extract the ultraviolet asymptotic structure relevant to the loop integral in the linear response. 
The adiabatic expansion is commonly used to organize ultraviolet divergences in the expectation value of the stress tensor for adiabatic renormalization. For unequal-time correlation functions, we must also account for the short-time behavior of the phase. This requires combining the adiabatic expansion with a large-momentum expansion that keeps the product of momentum and time separation finite.

The large-momentum asymptotic expansion of the mode functions does not coincide with the expansion in $\lambda$. 
In fact, we have
\begin{align}
\omega_p=p+\frac{a^2U}{2p}-\frac{a^4U^2}{8p^3}+O(p^{-5}).
\end{align}
Thus, while $\omega_p$ is leading-order~(LO) in the adiabatic expansion, it already contains higher-order terms in $p^{-1}$. The bookkeeping for large momentum is therefore different from that for $\lambda$.

The relevant ultraviolet structure is not determined by the large-momentum expansion of a single mode function alone. Because the response kernel involves unequal-time correlators, it also depends on the short-time limit. We must therefore take the large-momentum limit while keeping $p\Delta$ finite. This is crucial for evaluating the unequal-time correlation functions. Introducing
\begin{align}
\bar\tau=\frac{\tau+\tau'}2,\qquad \Delta=\tau-\tau',
\end{align}
we implement the joint limit by $p\to p/\eta$ and $\Delta\to\eta\Delta$, keeping $p\Delta$ finite. From Eq.~\eqref{frw23:wkb-ansatz}, the Green functions in the adiabatic approximation take the form
\begin{align}
G^R_p(\tau,\tau')
&=
\frac{\Theta(\tau-\tau')}{\sqrt{W_p(\tau)W_p(\tau')}}
\sin\!\left[\int_{\tau'}^\tau ds\,W_p(s)\right],
\\[4pt]
G^K_p(\tau,\tau')
&=
\frac{1}{\sqrt{W_p(\tau)W_p(\tau')}}
\cos\!\left[\int_{\tau'}^\tau ds\,W_p(s)\right].
\label{frw23:W-Green}
\end{align}
Since the leading-order contribution to the frequency in the phase is
\begin{align}
W=\mathcal O(p)=\mathcal O(\eta^{-1}),
\end{align}
we introduce a bookkeeping parameter $\eta=\mathcal O(p^{-1})=\mathcal O(\Delta)$. The phase of the Green function is then
\begin{align}
\int_{\tau_0}^{\tau}W(s)\,ds
-\int_{\tau_0}^{\tau-\Delta}W(s)\,ds
&=\int_{\tau-\Delta}^{\tau}W(s)\,ds
=\mathcal O(\Delta p).
\end{align}
Thus, after truncating the expansion of $W$ in $\eta$, an additional power of $\eta$ appears when the Green functions are evaluated.
We combine this phase factor with the amplitude to write $G^K$ and $G^R$.
The bookkeeping of $\eta$ is schematically given as follows. Let us first define
\begin{align}
A=\sum_{n=-1}^{\infty}a_n\eta^n,\qquad
B=\sum_{n=0}^{\infty}b_n\eta^n,
\end{align}
and consider
\begin{align}
\frac1A e^B
=\eta\frac{e^{b_0}}{a_{-1}}
\frac{1+b_1\eta+\cdots}
{1+\eta a_0/a_{-1}+\cdots}.
\end{align}
Green functions expanded in $\eta$ take this form, and to determine the terms through $\eta^n$, we must retain $a_{n-2}$ and $b_{n-1}$.
The required expansion orders are different for the amplitude and the phase.

We use $r$ to label the order of the ultraviolet expansion of $G^RG^K$ at fixed $p\Delta$ in Eq.~\eqref{frw23:response-kernel}. 
Since the radial weight is $O(p^6)$, the leading momentum dependence of the integrand at order $r$ can be written schematically as $p^{6-r}\mathcal G_r(p\Delta)$. 
Setting $x=p\Delta$ gives the short-time scaling
\begin{align}
\int_0^\infty dp\,p^{6-r}\mathcal G_r(p\Delta)
=\Delta^{r-7}\int_0^\infty dx\,x^{6-r}\mathcal G_r(x).
\label{frw23:r-counting}
\end{align}
Thus, the most singular short-time behavior at order $r$ is $\Delta^{r-7}$, and only terms with $r\leq6$ can produce local ultraviolet divergences in the subsequent time integral. At $r=6$, the leading singularity is $1/\Delta$. Terms with $r<6$ can also yield $1/\Delta$ contributions through subleading terms in the kernel and the short-time expansion of the external field. Terms with $r>6$ are integrable at $\Delta=0$ and finite.

To obtain $G^RG^K$ through $O(p^{-6})$, we need relative corrections through $O(p^{-4})$ to its leading term, which is $O(p^{-2})$. In the amplitude, corrections to $W_p$ are divided by its leading term, $p$. In the phase, they are integrated over a time interval of order $p^{-1}$. Thus, both require the expansion of $W_p$ through $O(p^{-3})$. From Eq.~\eqref{frw23:W-order}, we retain the adiabatic expansion through fourth order.

In what follows, we set $\lambda=1$ and write
\begin{align}
Q=a^2U-\frac{a''}a.
\label{frw23:q-definition}
\end{align}
From the fourth-order adiabatic solution above, the required large-momentum expansion is
\begin{align}
W_{p/\eta}^{[4]}
=\frac p\eta+\frac{\eta Q}{2p}
-\frac{\eta^3(Q^2+Q'')}{8p^3}+O(\eta^5).
\label{frw23:w-uv}
\end{align}
The corresponding mode function is
\begin{align}
u_{p/\eta}^{\rm A,[4]}(\tau,\tau_0)
\sim{}&\sqrt{\frac\eta{2p}}
\left[1-\frac{\eta^2Q}{4p^2}
+\frac{\eta^4(5Q^2+2Q'')}{32p^4}+O(\eta^6)\right]
\notag\\
&\times\exp\left[-\frac{ip(\tau-\tau_0)}\eta
-\frac{i\eta}{2p}\int_{\tau_0}^\tau Q\,ds
+\frac{i\eta^3}{8p^3}\int_{\tau_0}^\tau(Q^2+Q'')\,ds
+O(\eta^5)\right].
\label{frw23:mode-uv}
\end{align}
In the unequal-time Green functions, the phase is taken as the difference of its values at the two times, and the integration interval shrinks to $O(\eta)$; hence the remainder in Eq.~\eqref{frw23:w-uv} does not affect the Green functions retained through $O(\eta^5)$ below.

Substituting Eq.~\eqref{frw23:mode-uv} into Eqs.~\eqref{frw23:gr} and~\eqref{frw23:gk} and expanding the background functions about the central time, we obtain the retarded Green function (temporarily omitting the step function)
\begin{align}
&G^R_{p/\eta}\left(\bar\tau+\frac{\eta\Delta}2,
\bar\tau-\frac{\eta\Delta}2\right)
\notag\\
={}&\frac\eta p\sin(p\Delta)
+\frac{\eta^3Q}{2p^3}
\left[p\Delta\cos(p\Delta)-\sin(p\Delta)\right]
\notag\\
&+\frac{\eta^5}{p^5}\Biggl\{
\frac{Q^2}{8}\left[(3-p^2\Delta^2)\sin(p\Delta)
-3p\Delta\cos(p\Delta)\right]
\notag\\
&\hspace{13mm}+Q''\left[
\frac{2-p^2\Delta^2}{16}\sin(p\Delta)
+\frac{p^3\Delta^3-6p\Delta}{48}\cos(p\Delta)\right]
\Biggr\}+O(\eta^7),
\label{frw23:gr-uv}
\end{align}
and the Keldysh Green function
\begin{align}
&G^K_{p/\eta}\left(\bar\tau+\frac{\eta\Delta}2,
\bar\tau-\frac{\eta\Delta}2\right)
\notag\\
={}&\frac\eta p\cos(p\Delta)
-\frac{\eta^3Q}{2p^3}
\left[p\Delta\sin(p\Delta)+\cos(p\Delta)\right]
\notag\\
&+\frac{\eta^5}{p^5}\Biggl\{
\frac{Q^2}{8}\left[(3-p^2\Delta^2)\cos(p\Delta)
+3p\Delta\sin(p\Delta)\right]
\notag\\
&\hspace{13mm}+Q''\left[
\frac{2-p^2\Delta^2}{16}\cos(p\Delta)
-\frac{p^3\Delta^3-6p\Delta}{48}\sin(p\Delta)\right]
\Biggr\}+O(\eta^7).
\label{frw23:gk-uv}
\end{align}
Here $Q$ and its derivatives are all evaluated at $\bar\tau$.
\section{Renormalization of the linear response}\label{frw4:section}

We now use the adiabatic--ultraviolet expansion developed in the previous section to isolate the local ultraviolet part of the linear response. The bare response may be decomposed as
\begin{align}
T^{\rm lin,bare}_{ij}
&=\bigl(T^{\rm lin,bare}_{ij}-T^{\rm lin,adi}_{ij}\bigr)
+T^{\rm lin,adi}_{ij},
\end{align}
where $T^{\rm lin,adi}_{ij}$ includes both the adiabatic--ultraviolet expansion of the dynamical response and the local contact terms arising from the expansion of the action. We introduce counterterms so that
\begin{align}
T^{\rm lin,adi}_{ij}+T^{\rm lin,ct}_{ij}
=\text{finite part}.
\end{align}
The renormalized response may then be written as
\begin{align}
T^{\rm lin,ren}_{ij}
=\bigl(T^{\rm lin,bare}_{ij}-T^{\rm lin,adi}_{ij}\bigr)
+\bigl(T^{\rm lin,adi}_{ij}+T^{\rm lin,ct}_{ij}\bigr).
\label{frw4:subtraction}
\end{align}
The first term is ultraviolet finite, while the finite local part of the second term is fixed by the renormalization conditions. In what follows, we extract the local poles order by order and identify the required counterterms.

\subsection{General structure of the response kernel}

We compute the dynamical response in the adiabatic expansion:
\begin{align}
T^{\rm dyn,bare}_{ij}(\tau,\mathbf k)
&=\frac{\kappa}{a^2(\tau)}
\int_0^{\tau-\tau_0}d\Delta\int_0^\infty dp
\int_{|p-k|}^{p+k}dp'\,w(k,p,p')
\notag\\
&\quad\times G^R_{p'}(\tau,\tau-\Delta)
G^K_p(\tau,\tau-\Delta)
h_{ij}(\tau-\Delta,\mathbf k).
\label{frw4:response}
\end{align}
Here we first expand the Green functions around the center time $\bar\tau$, and only after the momentum integrations do we set $\bar\tau=\tau-\Delta/2$.
The results of the momentum integrations at each order may be summarized as
\begin{align}
T^{\rm dyn,adi}_{ij}(\tau,\mathbf k)
&=\frac1{a^2(\tau)}\int_0^{\tau-\tau_0}d\Delta
\left[A K(k\Delta)+B\frac{\partial K(k\Delta)}{\partial(k\Delta)}\right]
h_{ij}(\tau-\Delta,\mathbf k),
\label{frw4:ABkernel}
\end{align}
where $K(x)=j_2(x)/x^2$ and $j_2$ is the spherical Bessel function of order two. The kernel function $K$ has the same form as the damping kernel in collisionless kinetic theory~\cite{Rebhan:1994zw,Weinberg:2003ur}.
This representation isolates the short-distance singularities through the behavior of the coefficients $A$ and $B$ at small $\Delta$. An infrared expansion of the integrand of Eq.~\eqref{frw4:response} in powers of the external momentum $k$ would a priori destroy the nonlocal structure and obscure the $B$ coefficient. This is essentially a matter of uniform convergence, so the $(p,p')$ integrations should be carried out before any further expansion in $k$.
The coefficients are further expanded in $\eta$:
\begin{align}
A=\sum_r\eta^r A^{(r)},\qquad B=\sum_r\eta^r B^{(r)}.
\label{frw4:etaAB}
\end{align}
The relevant orders in the center-time expansion are $r=2,4,6$. In the coefficients below, the adiabatic contributions at the same order have been combined, and we set $\lambda=1$.

After the $p'$ integral, the $p$ integral reduces to
\begin{align}
\int_0^\infty dp\,p^n\sin^b(p\Delta)\cos^c(p\Delta),
\end{align}
where the background functions and their derivatives appear as coefficients. The adiabatic-ultraviolet expansion determines only the high-momentum asymptotic behavior, and extending it term by term down to $p=0$ in the integral may introduce spurious infrared divergences. We analytically continue these asymptotic integrals using dimensional regularization (DR), in order to separate the spurious infrared structure generated by the low-momentum extension from the local ultraviolet divergences to be extracted. We take the spatial dimension to be $D_{\rm space}=3-\epsilon$, corresponding to the spacetime dimension $d=4-\epsilon$, where $\epsilon\to0$ and $\mu$ is a renormalization scale with mass dimension. After reducing products of integer powers of sines and cosines to sums, we apply term by term
\begin{align}
\mu^\epsilon\int_0^\infty dp\,p^{n-\epsilon}\sin(p\Delta)
&=\frac{\mu^\epsilon\Gamma(n+1-\epsilon)}
{\Delta^{n+1-\epsilon}}
\sin\!\left[\frac\pi2(n+1-\epsilon)\right],
\label{frw4:sinDR}\\
\mu^\epsilon\int_0^\infty dp\,p^{n-\epsilon}\cos(p\Delta)
&=\frac{\mu^\epsilon\Gamma(n+1-\epsilon)}
{\Delta^{n+1-\epsilon}}
\cos\!\left[\frac\pi2(n+1-\epsilon)\right].
\label{frw4:cosDR}
\end{align}
Here $\Delta>0$; for higher harmonics, $\Delta$ is replaced by the corresponding multiple. The right-hand sides are obtained by analytic continuation of the integrals from their convergence regions. The factor $(\mu\Delta)^\epsilon$ in the time integrals below arises from the radial momentum integrals above, with the same regularization parameter.
Care is needed to retain all contributions at each order of the adiabatic-ultraviolet expansion, since an inconsistent truncation can leave spurious nonlocal divergences.
The genuine short-distance singularities appear as negative powers of $\Delta$, while terms containing positive powers of $\Delta$ are finite in the short-time integration. Scaleless power-law divergences vanish in DR. The required ultraviolet divergences are therefore determined by the singular terms in the coincidence limit $\Delta\to0$.

The singular behavior in the coincidence limit may be read off as follows. A typical time integral takes the form
\begin{align}
\int_0^{\tau-\tau_0}\frac{d\Delta}{\Delta}f(\Delta)
=\int_0^{\tau-\tau_0}\frac{d\Delta}{\Delta}
\left[f(0)+\Delta f'(0)+\cdots\right],
\end{align}
with $f'(0)=\partial_\Delta f(\Delta)|_{\Delta=0}$. The second term is finite in the limit $\Delta\to0$, while the integral of the first term is
\begin{align}
\int_0^{\tau-\tau_0}\frac{d\Delta}{\Delta}(\mu\Delta)^\epsilon
=\frac{[\mu(\tau-\tau_0)]^\epsilon}{\epsilon}.
\end{align}
Thus, the divergent part associated with $1/\Delta$ is expressed as
\begin{align}
\int_0^{\tau-\tau_0}\frac{d\Delta}{\Delta}f(\Delta)
=\frac{f(0)}\epsilon+(\text{finite part}).
\end{align}
Similarly, for $\Delta^{-2}$, we have
\begin{align}
\int_0^{\tau-\tau_0}\frac{d\Delta}{\Delta^2}f(\Delta)
&=\int_0^{\tau-\tau_0}\frac{d\Delta}{\Delta^2}
\left[f(0)+\Delta f'(0)+\frac{\Delta^2}{2!}f''(0)+\cdots\right]
\notag\\
&=f(0)\int_0^{\tau-\tau_0}\frac{d\Delta}{\Delta^2}
+f'(0)\int_0^{\tau-\tau_0}\frac{d\Delta}{\Delta}+(\text{finite part}).
\end{align}
In general, the short-time expansion gives
\begin{align}
\int_0^{\tau-\tau_0}\frac{d\Delta}{\Delta^n}f(\Delta)
=\sum_{m=0}^\infty\frac{f^{(m)}(0)}{m!}
\int_0^{\tau-\tau_0}\frac{d\Delta}{\Delta^{n-m}}.
\end{align}
The last integral is understood in dimensional regularization:
\begin{align}
\int_0^{\tau-\tau_0}\frac{d\Delta}{\Delta^{n-m}}(\mu\Delta)^\epsilon
=\frac{(\tau-\tau_0)^{1+m-n}}{1+m-n+\epsilon}[\mu(\tau-\tau_0)]^\epsilon.
\label{frw4:timeDR}
\end{align}
Thus, even in the present $3+1$ decomposition, one may employ a dimensional-regularization prescription directly analogous to the standard one in Poincar\'e-invariant theory.

By integration by parts, Eq.~\eqref{frw4:ABkernel} may be recast as
\begin{align}
a^2T^{\rm dyn,adi}_{ij}
&=F_1(\tau)K\bigl(k(\tau-\tau_0)\bigr)h_{ij}(\tau_0,\mathbf k)
+F_2(\tau,\partial_\tau,k)h_{ij}(\tau,\mathbf k)
\notag\\
&\quad+\int_{\tau_0}^\tau d\tau'\,
F_3(\tau,\tau')K\bigl(k(\tau-\tau')\bigr)
h'_{ij}(\tau',\mathbf k).
\end{align}
$F_1$ is a finite function of $\tau$, multiplied by the kinetic kernel $K\bigl(k(\tau-\tau_0)\bigr)$ and the initial perturbation $h_{ij}(\tau_0,\mathbf k)$. In the kinetic descriptions of the cited references, $K\bigl(k(\tau-\tau_0)\bigr)$ is a homogeneous solution of the linear response operator, and this term may be removed by an appropriate choice of initial condition~\cite{Rebhan:1994zw,Ota:2025rll}. In the present case, we can no longer maintain an exact kinetic description in terms of the Vlasov equation; this initial term should be treated within the full initial-value problem.

For the adiabatic vacuum used here, $F_1$ is ultraviolet finite. Choosing instead the instantaneous vacuum defined by diagonalizing the Hamiltonian at the initial time introduces a divergent contribution to $F_1$, which cannot be removed by bulk counterterms. We discuss this initial-state dependence in section~\ref{frw6:vacua}.
$F_2$ is the divergent local operator that we renormalize by the bulk counterterms. In the finite-temperature backgrounds, the finite local part of the response is exactly canceled by the contact response and the tadpole contribution from the dynamical background upon using the background equations~\cite{Ota:2025rll,Ota:2026yzp}. In the present FLRW setting, the finite local terms must be treated together with the background equations. Here we restrict our analysis to the ultraviolet divergences and the bulk counterterms required to cancel them.
In the above adiabatic expansion, $F_3$ is expressed by the finite parts of $B^{(2)}$, $B^{(4)}$, and $B^{(6)}$, together with higher orders in $r$. When nonadiabatic effects are included in the response, the kernel function will generally no longer take such a kinetic damping form. While the main goal of this paper is the renormalization of $F_2$, the determination of this finite contribution to the linear response in a general time-dependent background is the ultimate goal of this program, and will be discussed in more detail in future work.
\subsection{UV divergence order by order in \texorpdfstring{$r$}{r}}

The case $r=2$ corresponds to the ultraviolet leading term for a massless scalar field coupled to $h_{ij}$, whose local structure has already been obtained in the literature~\cite{Marunovic:2011zw}. The coefficients in Eq.~\eqref{frw4:ABkernel} at this order are
\begin{align}
A^{(2)}&=\frac{\kappa(10\Delta^2k^2-3)}{8\pi^2\Delta^5},
\\
B^{(2)}&=\frac{75\kappa k}{8\pi^2\Delta^4}.
\label{frw4:AB2}
\end{align}
Expanding the kernel functions in $\Delta$ and dropping $O(\Delta)$ terms, we find
\begin{align}
T^{\rm dyn,adi,(2)}_{ij}(\tau,\mathbf k)
&=\frac1{a^2}\int_0^{\tau-\tau_0} d\Delta
\left[-\frac{\kappa}{40\pi^2\Delta^5}
-\frac{\kappa k^2}{240\pi^2\Delta^3}
-\frac{\kappa k^4}{960\pi^2\Delta}\right]
\notag\\
&\hspace{25mm}\times h_{ij}(\tau-\Delta,\mathbf k)
+(\text{finite part}).
\label{frw4:kernel2}
\end{align}
Here $\Delta^{-n}$ is singular at $\Delta=0$. Using Eq.~\eqref{frw4:timeDR}, we obtain the pole
\begin{align}
T^{(2)}_{ij,\rm div}
=-\frac{\kappa}{960\pi^2\epsilon a^2}
\left(h''''_{ij}+2k^2h''_{ij}+k^4h_{ij}\right).
\label{frw4:pole2}
\end{align}
Since $\partial_\Delta^2=\partial_\tau^2$, this four-derivative structure is renormalized by the Weyl-squared counterterm.

\medskip
At $r=4$, both the mass correction in $W_{0,p}$ and the curvature correction in $W_{2,p}$ contribute, entering the coefficients through $Q=a^2U-a''/a$. We find
\begin{align}
A^{(4)}&=-\frac{5\kappa Q(\bar\tau)(3\Delta^2k^2+1)}
{16\pi^2\Delta^3},
\\
B^{(4)}&=-\frac{85\kappa k Q(\bar\tau)}{16\pi^2\Delta^2},
\label{frw4:AB4}
\end{align}
with
\begin{align}
\bar\tau=\tau-\frac{\eta\Delta}{2}.
\label{frw4:midpoint}
\end{align}
In terms of the observation time, we then have
\begin{align}
T^{\rm dyn,adi,(4)}_{ij}(\tau,\mathbf k)
&=\frac1{a^2}\int_0^{\tau-\tau_0} d\Delta
\left[-\frac{\kappa}{48\pi^2\Delta^3}
\left(a^2U-\frac{a''}a\right)
-\frac{\kappa k^2}{96\pi^2\Delta}
\left(a^2U-\frac{a''}a\right)\right]
\notag\\
&\hspace{25mm}\times h_{ij}(\tau-\Delta,\mathbf k)
+(\text{finite part}).
\label{frw4:kernel4}
\end{align}
We use the same treatment as for $r=2$ and obtain
\begin{align}
T^{(4)}_{ij,\rm div}
=-\frac{\kappa}{96\pi^2\epsilon a^2}
\left(a^2U-\frac{a''}a\right)
\left(h''_{ij}+k^2h_{ij}\right).
\label{frw4:pole4}
\end{align}
This $\partial_\Delta^2+k^2=\partial_\tau^2+k^2=\Box $ structure is renormalized jointly by the counterterms for the time-dependent Ricci-scalar term and the Ricci-scalar-squared term.
We first expand the Green functions around the center time $\bar\tau=(\tau+\tau')/2$, and then write $\bar\tau=\tau-\Delta/2$ after the momentum integrations. The leading $r=2$ term depends only on the time difference $\Delta$; because the $r=4$ coefficient depends explicitly on the center time, rewriting it as a function of the observation time and the time difference generates an additional local term proportional to $h'_{ij}$. Its divergent contribution is
\begin{align}
T^{(5)}_{ij,\rm div}
=-\frac{\kappa}{96\pi^2\epsilon a^2}
\left(2aa'U+a^2U'+\frac{a'a''}{a^2}-\frac{a'''}a\right)h'_{ij},
\label{frw4:pole5}
\end{align}
where we used $\partial_\Delta h_{ij}(\tau-\Delta)=-h'_{ij}(\tau-\Delta)$.
This structure proportional to $h'_{ij}$ arises from the variations of the time-dependent Ricci-scalar term and the Ricci-scalar-squared term.
A term proportional to $h^{ij}h'_{ij}$ in the effective action is a total derivative when its coefficient is constant. Here the coefficient depends on time through the background geometry and the counterterm coefficients, so such a term cannot in general be discarded as a boundary term.

\medskip
At $r=6$, both leading-order and higher-order adiabatic corrections contribute. If one keeps only the $\lambda^2$ part, an apparent nonlocal divergence arises from the Ci functions. This is exactly canceled once the full contribution at this order is included, so in the final result we set $\lambda=1$ and obtain
\begin{align}
A^{(6)}
&=\frac{5\kappa Q(\bar\tau)^2(\Delta^2k^2-3)}{32\pi^2\Delta}
+\frac{5\kappa Q''(\bar\tau)(\Delta^2k^2-9)}{384\pi^2\Delta},
\\
B^{(6)}
&=\frac{5\kappa k Q(\bar\tau)^2}{32\pi^2}
-\frac{5\kappa k Q''(\bar\tau)}{384\pi^2}.
\label{frw4:AB6}
\end{align}
The leading terms in $B^{(6)}$ have no short-time singularity. $A^{(6)}$ produces a local term proportional to $h_{ij}$, which, together with the zero-derivative terms arising from the variations of the Ricci-scalar and curvature-squared counterterms, enters the matching of the cosmological-constant term.
Finally, by expanding $K(k\Delta)$ in $\Delta$ and including the quadratic Taylor term generated by rewriting the $r=4$ coefficient from the center time to the observation time, we obtain
\begin{align}
T^{\rm dyn,adi,(6)}_{ij}(\tau,\mathbf k)
&=-\frac{\kappa}{96\pi^2a^2}\int_0^{\tau-\tau_0}\frac{d\Delta}{\Delta}
\biggl[3a^4U^2+2(a')^2U+4aa'U'-4aa''U+a^2U''
\notag\\[-1mm]
&\hspace{27mm}
-\frac{2(a')^2a''}{a^3}
+\frac{4(a'')^2+2a'a'''}{a^2}-\frac{a''''}a\biggr]
\notag\\
&\hspace{25mm}\times h_{ij}(\tau-\Delta,\mathbf k)
+(\text{finite part}).
\label{frw4:kernel6}
\end{align}
Therefore,
\begin{align}
T^{(6)}_{ij,\rm div}
&=-\frac{\kappa}{96\pi^2\epsilon a^2}
\biggl[3a^4U^2+2(a')^2U+4aa'U'-4aa''U+a^2U''
\notag\\[-1mm]
&\hspace{30mm}
-\frac{2(a')^2a''}{a^3}
+\frac{4(a'')^2+2a'a'''}{a^2}-\frac{a''''}a\biggr]h_{ij}.
\label{frw4:pole6}
\end{align}
Together with the ultraviolet power counting in the previous section, this shows that higher orders introduce no further nonintegrable singularities. Thus, the divergent local structure relevant for ultraviolet renormalization is exhausted by the three classes of operators $\Box ^2h_{ij}$, $(\Box ,\partial_\tau)h_{ij}$, and $h_{ij}$.
The contact contribution in Eq.~\eqref{frw23:full-response} is
\begin{align}
T^{\rm cont}_{ij}
=-\frac{\kappa}{a^2}h_{ij}
\langle 0|\mathcal L^{(0)}_{\psi}|0\rangle.
\label{frw4:contact}
\end{align}
Evaluating it using the mode functions of the field variable $\psi$, we have
\begin{align}
\langle 0|\mathcal L^{(0)}_{\psi}|0\rangle
=\frac1{4\pi^2}\int_0^\infty dp\,p^2
\left[|u_p'-\mathcal H u_p|^2-(p^2+a^2U)|u_p|^2\right].
\label{frw4:contactmodes}
\end{align}
Substituting the adiabatic expansion from the previous section and retaining the $1/p$ terms, we obtain
\begin{align}
T^{\rm cont}_{ij,\rm div}
&=\frac{\kappa}{32\pi^2\epsilon a^2}
\biggl[a^2U''+2aa'U'
-\frac{6(a')^2a''}{a^3}
\notag\\[-1mm]
&\hspace{30mm}
+\frac{3(a'')^2+4a'a'''}{a^2}-\frac{a''''}a\biggr]h_{ij}.
\label{frw4:contactpole}
\end{align}

\subsection{Renormalization}

We use the covariant local counterterms
\begin{align}
S^{\rm ct}
=\int d^4x\sqrt{-g}
\left[c_1+c_2R+c_3C_{\mu\nu\rho\sigma}C^{\mu\nu\rho\sigma}
+c_5R^2\right].
\label{frw4:ctbasis}
\end{align}
The four-dimensional Gauss--Bonnet term with a constant coefficient does not change the local variation in the interior, and covariant total derivatives contribute only endpoint terms. They are therefore omitted from Eq.~\eqref{frw4:ctbasis}. The counterterm stress tensor is given by
\begin{align}
T^{\rm ct,lin}_{ij}
=-\frac2{\kappa a^2}
\frac{\delta S^{\rm ct,(2)}}{\delta h^{ij}}.
\label{frw4:ctvariation}
\end{align}

We determine the counterterm coefficients by matching the operators in the linear response in descending order of the number of derivatives acting on $h_{ij}$.
The four-dimensional Weyl-squared action is conformally invariant, and its second-order expansion is
\begin{align}
S^{c_3,(2)}
=\frac{\kappa^2c_3}{2}\int d^4x
\left(h''_{ij}-\nabla^2h_{ij}\right)^2,
\end{align}
where total derivatives have been omitted; these terms do not change the local variation in the interior of the interval. Therefore,
\begin{align}
T^{c_3,\rm lin}_{ij}
=-\frac{2\kappa c_3}{a^2}
\left(h''''_{ij}+2k^2h''_{ij}+k^4h_{ij}\right).
\label{frw4:weylvariation}
\end{align}
First matching the four-derivative structure in Eq.~\eqref{frw4:pole2}, we obtain
\begin{align}
c_3=-\frac1{1920\pi^2\epsilon}.
\label{frw4:c3}
\end{align}

On the FLRW background, $R_0=6a''/a^3$. Although the TT perturbation satisfies $R^{(1)}=0$, the second-order expansion of $R^2$ contains $2R_0R^{(2)}$, so it participates in the remaining matching. In the inverse-metric perturbation parametrization, the second-order Ricci scalar, after dropping spatial total derivatives, is
\begin{align}
R^{(2)}
=\frac{\kappa^2}{a^2}
\left[h^{ij}h''_{ij}+\frac54h'^{ij}h'_{ij}
-\frac14\partial_\ell h^{ij}\partial_\ell h_{ij}
+3\mathcal H h^{ij}h'_{ij}\right].
\label{frw4:R2}
\end{align}
Substituting this together with the determinant expansion gives
\begin{align}
S^{c_1,(2)}
&=\frac{\kappa^2}{4}\int d^4x\,a^4c_1h^{ij}h_{ij},
\label{frw4:c1action}\\
S^{c_2,(2)}
&=\int d^4x\,a^4c_2
\left[R^{(2)}+\frac{\kappa^2R_0}{4}h^{ij}h_{ij}\right],
\label{frw4:c2action}\\
S^{c_5,(2)}
&=c_5\int d^4x\,a^4
\left[2R_0R^{(2)}+\frac{\kappa^2R_0^2}{4}h^{ij}h_{ij}\right].
\label{frw4:c5action}
\end{align}
Varying according to Eq.~\eqref{frw4:ctvariation}, we obtain, respectively,
\begin{align}
T^{c_1,\rm lin}_{ij}
&=-\kappa a^2c_1h_{ij},
\label{frw4:ct1}\\
T^{c_2,\rm lin}_{ij}
&=\frac\kappa{a^2}\biggl\{
a^2c_2(h''_{ij}+k^2h_{ij})
+(2aa'c_2+a^2c_2')h'_{ij}
\notag\\
&\hspace{12mm}
+\left[2(a')^2c_2-4aa''c_2-2aa'c_2'-2a^2c_2''\right]h_{ij}
\biggr\},
\label{frw4:ct2}\\
T^{c_5,\rm lin}_{ij}
&=\frac{12\kappa c_5}{a^2}\biggl\{
\frac{a''}a(h''_{ij}+k^2h_{ij})
+\left(\frac{a'''}a-\frac{a'a''}{a^2}\right)h'_{ij}
\notag\\
&\hspace{12mm}
+\left[-\frac{2a''''}a
+\frac{10a'a'''+5(a'')^2}{a^2}
-\frac{16(a')^2a''}{a^3}\right]h_{ij}\biggr\}.
\label{frw4:ct5}
\end{align}

Next, matching the $r=4$ structure $h''_{ij}+k^2h_{ij}$ requires
\begin{align}
a^2c_2+12c_5\frac{a''}a
=\frac1{96\pi^2\epsilon}
\left(a^2U-\frac{a''}a\right).
\label{frw4:kinmatching}
\end{align}
This relation must hold for arbitrary smooth $a$ and $U$. Comparing the independent mass and curvature structures separately, we obtain
\begin{align}
c_2=\frac{U}{96\pi^2\epsilon},
\qquad
c_5=-\frac1{1152\pi^2\epsilon}.
\label{frw4:c2c5}
\end{align}
Substituting into the $h_{ij}$ terms of Eqs.~\eqref{frw4:ct2} and \eqref{frw4:ct5}, and adding Eqs.~\eqref{frw4:pole6} and \eqref{frw4:contactpole}, leaves the zero-derivative structure
\begin{align}
T^{(6)}_{ij,\rm div}+T^{\rm cont}_{ij,\rm div}
&+\left.T^{c_2,\rm lin}_{ij}\right|_{h_{ij}}
+\left.T^{c_5,\rm lin}_{ij}\right|_{h_{ij}}
=-\frac{\kappa a^2U^2}{32\pi^2\epsilon}h_{ij}.
\end{align}
Equation~\eqref{frw4:ct1} therefore fixes
\begin{align}
c_1=-\frac{U^2}{32\pi^2\epsilon}.
\label{frw4:c1}
\end{align}
Finally, differentiating Eq.~\eqref{frw4:kinmatching} with respect to conformal time gives
\begin{align}
(a^2c_2)'+12c_5\left(\frac{a''}a\right)'
=\frac1{96\pi^2\epsilon}
\left(2aa'U+a^2U'+\frac{a'a''}{a^2}-\frac{a'''}a\right).
\label{frw4:r5check}
\end{align}
This is precisely the coefficient of $h'_{ij}$ in Eqs.~\eqref{frw4:ct2} and \eqref{frw4:ct5}, which automatically cancels Eq.~\eqref{frw4:pole5} without introducing a new renormalization coefficient.

Thus, the same set of coefficients $c_1,c_2,c_3,c_5$ removes all local poles. $c_1$ and $c_2$ depend on $U$ and need not contain $U'$ or $U''$; the time derivatives of the mass appearing in the response arise from the variation of the time-dependent counterterm coefficients, while the curvature derivatives arise from the variation of the curvature counterterms.

The four coefficients above agree with the local counterterms in the DeWitt--Schwinger expansion~\cite{Vassilevich:2003xt}. In particular, $c_3$ and $c_5$ also agree, respectively, with the curvature-squared counterterms given in Ref.~\cite{Marunovic:2011zw}, expressed in the basis used here. A simple example of Pauli--Villars regularization for the tadpole is given at the end of section~\ref{frw6:resonance}.

\section{Consistency with the renormalization of tadpole}\label{frw5:sec}

Although the main focus of this paper is the renormalization of the gravitational-wave linear response, the same counterterms should also remove the ultraviolet divergences of the tadpole stress tensor in the unperturbed background. To check this consistency, we evaluate the energy density and pressure in the FLRW background and independently determine the required counterterm coefficients.

Here we evaluate the physical stress tensor of the original field $\chi$, defined by varying $S_\chi$ with respect to the metric; $\psi=a\chi$ is used only for the calculation, and the corresponding scale factors and $\mathcal H$ terms must be retained. The scalar stress tensor is
\begin{equation}
 T_{\mu\nu}^{\mathrm{tad}}
 =\partial_\mu\chi\partial_\nu\chi
 -\frac12g_{\mu\nu}
 \left(g^{\rho\sigma}\partial_\rho\chi\partial_\sigma\chi+U\chi^2\right).
 \label{frw5:stress}
\end{equation}
In conformal coordinates, the physical energy density and pressure are
$\rho^{\mathrm{tad}}=a^{-2}T_{00}^{\mathrm{tad}}$ and
$P^{\mathrm{tad}}=(3a^2)^{-1}\delta^{ij}T_{ij}^{\mathrm{tad}}$, namely
\begin{align}
 \rho^{\mathrm{tad}}
 &=\frac1{2a^2}\chi'^2
 +\frac1{2a^2}(\partial_i\chi)^2+\frac12U\chi^2,
 \notag\\
 P^{\mathrm{tad}}
 &=\frac1{2a^2}\chi'^2
 -\frac1{6a^2}(\partial_i\chi)^2-\frac12U\chi^2.
 \label{frw5:rhoPoperator}
\end{align}
We therefore need $\langle\chi'^2\rangle$, $\langle(\partial_i\chi)^2\rangle$, and $\langle\chi^2\rangle$.
Using $\psi=a\chi$ and $\pi_\psi=\psi'-\mathcal H\psi$, we have
$\chi'=a^{-1}\pi_\psi=a^{-1}(\psi'-\mathcal H\psi)$. We have
\begin{align}
 \chi'(\tau,\vec x)
 &=\frac1a\int\frac{d^3p}{(2\pi)^3}e^{i\vec p\cdot\vec x}
 \left[(u_p'-\mathcal H u_p)a_{\vec p}
 +(u_p^{*\prime}-\mathcal H u_p^*)a^\dagger_{-\vec p}\right],
 \notag\\
 \partial_i\chi(\tau,\vec x)
 &=\frac1a\int\frac{d^3p}{(2\pi)^3}
 ip_i e^{i\vec p\cdot\vec x}
 \left[u_p(\tau)a_{\vec p}
 +u_p^*(\tau)a^\dagger_{-\vec p}\right].
 \label{frw5:derivative-modes}
\end{align}
Here $u_p$ is the mode function in the adiabatic expansion.
The creation and annihilation operators are defined by the adiabatic vacuum at the initial time $\tau_0$, i.e., $a_{\vec p}|0\rangle=0$. Then
\begin{align}
 \langle0|\chi'^2|0\rangle
 &=\frac1{a^2}\int\frac{d^3p}{(2\pi)^3}|u_p'-\mathcal H u_p|^2,
 \notag\\
 \langle0|(\partial_i\chi)^2|0\rangle
 &=\frac1{a^2}\int\frac{d^3p}{(2\pi)^3}p^2|u_p|^2,
 \notag\\
 \langle0|\chi^2|0\rangle
 &=\frac1{a^2}\int\frac{d^3p}{(2\pi)^3}|u_p|^2.
 \label{frw5:expectation-modes}
\end{align}
\Needspace{8\baselineskip}
Substituting these expectation values into eq.~\eqref{frw5:rhoPoperator} and performing the angular integral, we obtain
\begin{align}
 \langle\rho^{\mathrm{tad}}\rangle
 &=\frac1{4\pi^2a^4}\int_0^\infty dp\,p^2
 \left[|u_p'-\mathcal H u_p|^2+(p^2+a^2U)|u_p|^2\right],
 \label{frw5:rhomode}
 \\
 \langle P^{\mathrm{tad}}\rangle
 &=\frac1{4\pi^2a^4}\int_0^\infty dp\,p^2
 \left[|u_p'-\mathcal H u_p|^2
 -\left(\frac{p^2}{3}+a^2U\right)|u_p|^2\right].
 \label{frw5:Pmode}
\end{align}

We extract the high-momentum terms using the fourth-order adiabatic expansion. With $Q$ defined in eq.~\eqref{frw23:q-definition}, the required asymptotic forms of the mode functions are
\begin{align}
 |u_p|^2
 &\sim\frac1{2p}-\frac{Q}{4p^3}
 +\frac{3Q^2+Q''}{16p^5}+O(p^{-7}),
 \\
 |u_p'|^2
 &\sim\frac p2+\frac{Q}{4p}
 -\frac{Q^2+Q''}{16p^3}+O(p^{-5}),
 \\
 \operatorname{Re}(u_p'u_p^*)
 &\sim-\frac{Q'}{8p^3}+O(p^{-5}).
 \label{frw5:modeUV}
\end{align}
Substituting these into eqs.~\eqref{frw5:rhomode} and \eqref{frw5:Pmode}, we extract the $p^{-1}$ term in the radial integral using the dimensional-regularization prescription described above. For a fixed high-momentum lower limit $p_0>0$,
\begin{equation}
 \mu^\epsilon\int_{p_0}^\infty dp\,p^{-1-\epsilon}
 =\frac1\epsilon+O(\epsilon^0).
 \label{frw5:radialpole}
\end{equation}
The resulting poles of the energy density and pressure are
\begin{align}
 \rho_{\mathrm{div}}
 &=\frac1{32\pi^2\epsilon}
 \left[
 -U^2+\frac{2U(a')^2}{a^4}
 +\frac{2a'U'}{a^3}
 +\frac{4(a')^2a''}{a^7}
 +\frac{(a'')^2-2a'a'''}{a^6}
 \right],
 \label{frw5:rhopole}
 \\
 P_{\mathrm{div}}
 &=\frac1{96\pi^2\epsilon}
 \left[
 3U^2+\frac{2U(a')^2}{a^4}
 -\frac{2a'U'+4Ua''}{a^3}
 -\frac{2U''}{a^2}
 \right.\nonumber\\[-2pt]
 &\hspace{26mm}\left.
 +\frac{16(a')^2a''}{a^7}
 -\frac{5(a'')^2+10a'a'''}{a^6}
 +\frac{2a''''}{a^5}
 \right].
 \label{frw5:Ppole}
\end{align}

At leading adiabatic order, this reproduces the equation of state of a cosmological constant, $P=-\rho$, though with time-dependent coefficients.

The Weyl tensor vanishes in the FLRW background, so $c_3$ does not contribute to the background tadpole; the Gauss--Bonnet term does not affect the local variation in the four-dimensional bulk. The stress tensors of the remaining counterterms are
\begin{align}
 T^{c_1}_{\mu\nu}&=c_1g_{\mu\nu},
 \\
 T^{c_2}_{\mu\nu}
 &=-2\left[c_2G_{\mu\nu}
 +(-g_{\mu\nu}\Box_g-\nabla_\mu\nabla_\nu)c_2\right],
 \label{frw5:c2stress}
 \\
 T^{c_5}_{\mu\nu}
 &=-2c_5\left[2RR_{\mu\nu}-\frac12g_{\mu\nu}R^2
 +2(-g_{\mu\nu}\Box_g-\nabla_\mu\nabla_\nu)R\right].
 \label{frw5:c5stress}
\end{align}
Here $G_{\mu\nu}$ is the Einstein tensor. We denote the covariant d'Alembertian by $\Box_g=-g^{\rho\sigma}\nabla_\rho\nabla_\sigma$. For a scalar $f$ on the FLRW background, $\Box_g f=a^{-2}(\Box f+2\mathcal H f')$, where $\Box=\partial_\tau^2-\nabla^2$ as defined in section~\ref{frw23:setup}. We leave the functional forms of $c_1(U)$ and $c_2(U)$ unspecified and treat $c_5$ as an undetermined constant. In a general FLRW background, the counterterm contributions are
\begin{align}
 \rho^{\mathrm{ct}}
 &=-c_1-\frac{6(a')^2}{a^4}c_2-\frac{6a'}{a^3}c_2'
 +c_5\left[
 \frac{144(a')^2a''}{a^7}
 +\frac{36(a'')^2-72a'a'''}{a^6}
 \right],
 \label{frw5:rhoct}
 \\
 P^{\mathrm{ct}}
 &=c_1+\left(\frac{4a''}{a^3}-\frac{2(a')^2}{a^4}\right)c_2
 +\frac{2c_2''}{a^2}+\frac{2a'c_2'}{a^3}
 \nonumber\\[-2pt]
 &\quad+c_5\left[
 \frac{192(a')^2a''}{a^7}
 -\frac{60(a'')^2+120a'a'''}{a^6}
 +\frac{24a''''}{a^5}
 \right].
 \label{frw5:Pct}
\end{align}
Since $c_2$ depends on time, the stress tensor of the Ricci-scalar term contains $c_2'$ and $c_2''$. In the FLRW background, $G_{\mu\nu}$ is generally nonzero, so this counterterm contributes to both the energy density and pressure, as shown above.
Since $a$ and $U$ are arbitrary smooth background functions, the condition
$\rho_{\mathrm{div}}+\rho^{\mathrm{ct}}=0$ must hold for each independent background structure. Comparing successively the terms proportional to $a'a'''/a^6$, $(a')^2/a^4$, and those with no background derivatives, we obtain
\begin{align}
 -\frac1{16\pi^2\epsilon}-72c_5&=0,
 \\
 \frac{U}{16\pi^2\epsilon}-6c_2&=0,
 \\
 -\frac{U^2}{32\pi^2\epsilon}-c_1&=0.
 \label{frw5:match}
\end{align}
Thus the required pole coefficients are
\begin{equation}
 c_1=-\frac{U^2}{32\pi^2\epsilon},\qquad
 c_2=\frac{U}{96\pi^2\epsilon},\qquad
 c_5=-\frac1{1152\pi^2\epsilon}.
 \label{frw5:ctcoeffs}
\end{equation}
The resulting $c_2'=U'/(96\pi^2\epsilon)$ also cancels the $a'U'/a^3$ term in the energy density, while $c_5$ cancels the remaining purely geometric terms. Substituting the same coefficients into the pressure independently gives
\begin{equation}
 \rho_{\mathrm{div}}+\rho^{\mathrm{ct}}=0,
 \qquad
 P_{\mathrm{div}}+P^{\mathrm{ct}}=0.
 \label{frw5:cancellation}
\end{equation}
The three coefficients determined from the tadpole agree with those obtained from the complete linear response. Since the Weyl tensor vanishes in the FLRW background, $c_3$ can only be determined from the four-derivative structure of the linear response and cannot be independently checked by the background tadpole.

\section{An example: Parametric resonance}\label{frw6:resonance}

With future applications to preheating after inflation in mind, we apply the results of the preceding sections to a scalar field whose effective mass is generated by a coherently oscillating background.
On time scales over which cosmic expansion and the variation of the oscillation amplitude can be neglected, the preheating mode equation reduces to the Mathieu equation in flat spacetime, and we give a calculation in this limit.
This approximation is insufficient for quantitative predictions, since the resonance bands shift as the universe expands and the linear response is dominant on scales near the Hubble radius.
Our present aim, however, is to understand the qualitative features of the gravitational-wave response in a resonant medium.
For this purpose, the simplified model is convenient because its mode functions can be evaluated using numerical libraries for Mathieu functions.
In what follows we identify the divergent structure of the Mathieu oscillator and discuss the adiabatic-ultraviolet asymptotics using our results.
We also illustrate Pauli--Villars regularization for the tadpole in the flat-background Mathieu model.

\subsection{Coherently oscillating background}

Let us consider a Mathieu-like oscillator in a flat spacetime: 
\begin{align}
U(t)=4m^2q\sin^2(mt),
\label{frw5:pv-example-mass}
\end{align}
with $a=1$ and $t=\tau$.
This type of oscillator can arise during preheating after inflation in the model~\cite{Kofman:1997yn,Mukhanov:2005sc}
\begin{align}   
    S=\int d^4x\sqrt{-g}\biggl[
&\frac{2}{\kappa^2}R
-\frac12g^{\mu\nu}\partial_\mu\phi\,\partial_\nu\phi
-\frac12m^2\phi^2\notag\\
&-\frac12g^{\mu\nu}\partial_\mu\chi\,\partial_\nu\chi
-\frac12g^2\phi^2\chi^2
\biggr].
\end{align}
After inflation, the inflaton condensate oscillates about the minimum of its quadratic potential as
\begin{align}
    \phi(t)=\Phi(t)\sin(mt).
\end{align}
When the backreaction of particle production can be neglected, the oscillation-averaged background evolves approximately as a matter-dominated universe.
The amplitude satisfies $\dot\Phi/\Phi\simeq-3H/2$, where $H=\dot a/a$. It therefore varies slowly compared with the oscillation when $1\ll m/H \sim mt$.
Over a time interval much shorter than $H^{-1}$, the fractional changes in the scale factor and the oscillation amplitude are small. We neglect these changes and use a flat-background model with $a=1$ and constant $\Phi$. We then introduce
\begin{align}
    q&=\frac{g^2\Phi^2}{4m^2},
    \\
    A_p&=\frac{p^2}{m^2}+2q.
\end{align}
Then, the equation of motion for $\chi$ in Fourier space becomes the Mathieu equation 
\begin{align}
\frac{d^2\chi_p}{dt^2}
+m^2\bigl[A_p-2q\cos(2mt) \bigr]\chi_p=0.
\label{frw6:res-mathieu}
\end{align}
This is the Mathieu equation familiar from parametric resonance~\cite{Kofman:1994rk,Kofman:1997yn}.
Narrow resonance corresponds to $q\ll1$, with the $n$th resonance band near $A_p\simeq n^2$, with an integer $n$.
Long-term growth near the center of the first band occurs on the time scale $\Delta\theta=O(q^{-1})$, or $\Delta t=O((qm)^{-1})$, where $\theta=mt$. Broad resonance generally corresponds to $q\gg1$, where the relevant low-momentum modes can undergo nonadiabatic particle production near zero crossings of the background field~\cite{Kofman:1997yn}. In FLRW, $A_p$ and $q$ vary with time, and modes can cross resonance bands as the background evolves.
However, this flat spacetime model would be enough to capture the resonance feature in the gravitational wave response.
We focus on the ultraviolet structure as the internal momentum $p\to\infty$ for fixed smooth background and model parameters. In this limit $A_p\gg q$.

We can canonically quantize the Mathieu oscillator as usual.
The mode functions can be obtained using the procedure presented in Sec.~\ref{adiabaticsection}, i.e., one may find a mode function $u_{p}(t,t_0)$ in the adiabatic expansion and write 
\begin{align}
  \chi (t,\vec x) &=\int\frac{d^3p}{(2\pi)^3}e^{i\vec p\cdot\vec x}
\left[a_{\vec p}(t_0)u_p(t,t_0)
+a^\dagger_{-\vec p}(t_0)u_p^*(t,t_0)\right],
\end{align}
with the adiabatic vacuum defined at $t=t_0$: 
\begin{align}
    a_{\vec p}(t_0) |0\rangle = 0.
\end{align}
The remaining calculation follows from the general results of the preceding sections with the mass in Eq.~\eqref{frw5:pv-example-mass}.

\subsection{Ultraviolet divergences of the linear response}

The effective mass squared is given by Eq.~\eqref{frw5:pv-example-mass} and its time derivatives are
\begin{align}
\dot U&=4m^3q\sin(2mt),
\label{frw5:pv-example-massprime}\\
\ddot U&=8m^4q\cos(2mt).
\label{frw5:pv-example-masssecond}
\end{align}
Substituting eqs.~\eqref{frw5:pv-example-mass}, \eqref{frw5:pv-example-massprime}, and \eqref{frw5:pv-example-masssecond} into eqs.~\eqref{frw4:pole2}, \eqref{frw4:pole4}, \eqref{frw4:pole5}, and~\eqref{frw4:pole6}, and using $\Box=\partial_t^2+k^2$ for spatial Fourier modes, we obtain
\begin{align}
T^{(2)}_{ij,\mathrm{div}}
&=-\frac{\kappa}{960\pi^2\epsilon}\Box^2h_{ij},
\label{frw6:res-r2}\\
T^{(4)}_{ij,\mathrm{div}}
&=-\frac{\kappa m^2q\sin^2(mt)}{24\pi^2\epsilon}\Box h_{ij},
\label{frw6:res-r4}\\
T^{(5)}_{ij,\mathrm{div}}
&=-\frac{\kappa m^3q\sin(2mt)}{24\pi^2\epsilon}\dot h_{ij},
\label{frw6:res-r5}
\\
T^{\mathrm{dyn},(6)}_{ij,\mathrm{div}}
&=-\frac{\kappa m^4}{\pi^2\epsilon}
\left[\frac{q^2}{2}\sin^4(mt)+\frac{q}{12}\cos(2mt)\right]h_{ij},
\label{frw6:res-r6}
\\
T^{\mathrm{cont}}_{ij,\mathrm{div}}
&=\frac{\kappa m^4q\cos(2mt)}{4\pi^2\epsilon}h_{ij}.
\label{frw6:res-contact}
\end{align}
The local divergent part of the complete linear response is the sum of eqs.~\eqref{frw6:res-r2}--\eqref{frw6:res-contact}.
In this way, we can separate the ultraviolet contribution by using the adiabatic-ultraviolet expansion developed in this work.
Our systematic expansion scheme will be useful for evaluating the remaining finite contribution in our future work.

\subsection{Adiabatic versus instantaneous vacua}
\label{frw6:vacua}

We now compare the adiabatic vacuum used above with the instantaneous vacuum defined by diagonalizing the Hamiltonian at the initial time.
These vacuum conditions are not equivalent and they yield different ultraviolet structures. 

The instantaneous vacuum is specified by a mode-function basis that diagonalizes the Hamiltonian at the initial time $t_0=0$.
Equivalently, the mode functions and the instantaneous annihilation operator satisfy
\begin{align}
a_{\vec p}^{\rm inst}(0)|0_{\rm inst}\rangle=0,
\qquad
u_p^{\rm inst}(0)=\frac1{\sqrt{2p}},
\qquad
v_p^{\rm inst}(0)=-i\sqrt{\frac p2}.
\label{frw6:inst-initial}
\end{align}
Here and below, a dot denotes a derivative with respect to $t$. The mode satisfying these conditions can be written in terms of the Mathieu cosine $C_p(mt)$ and the Mathieu sine $S_p(mt)$. We normalize them by
\begin{align}
C_p(0)=1,\qquad \dot C_p(0)=0,\qquad
S_p(0)=0,\qquad \dot S_p(0)=m.
\label{frw6:mathieu-normalization}
\end{align}
The instantaneous-vacuum mode is
\begin{align}
u_p^{\rm inst}(t)
=\frac1{\sqrt{2p}}
\left[C_p(mt)-i\frac p m S_p(mt)\right].
\label{frw6:mathieu-instantaneous}
\end{align}
The identity $C_p(mt)\dot S_p(mt)-\dot C_p(mt)S_p(mt)=m$ ensures the Wronskian condition
$u_p^{\rm inst}v_p^{{\rm inst}*}-v_p^{\rm inst}u_p^{{\rm inst}*}=i$.

The positive-frequency adiabatic mode has a different normalization at $t=0$. In the flat background, its frequency has the ultraviolet expansion
\begin{align}
W_p(t)
=p+\frac{U}{2p}-\frac{U^2+\ddot U}{8p^3}+O(p^{-5}).
\label{frw6:vacuum-W-uv}
\end{align}
Although $U(0)=\dot U(0)=0$, we have $\ddot U(0)=8qm^4$, and hence
\begin{align}
W_{p0}\equiv W_p(0)
=p-\frac{qm^4}{p^3}+O(p^{-5}),
\qquad \dot W_p(0)=0.
\label{frw6:vacuum-W0}
\end{align}
For the comparison, choose a positive $W_{p0}$ that agrees with the fourth-order adiabatic frequency in the ultraviolet. The mode with the corresponding adiabatic initial values, evolved with the Mathieu equation, is
\begin{align}
u_p^{\rm A}(t)
=\frac1{\sqrt{2W_{p0}}}
\left[C_p(mt)-i\frac{W_{p0}}mS_p(mt)\right].
\label{frw6:mathieu-adiabatic}
\end{align}
This construction uses the adiabatic expansion to specify the initial values and the Mathieu equation for the subsequent evolution. In particular,
\begin{align}
u_p^{\rm A}(0)
=\frac1{\sqrt{2p}}
\left[1+\frac{qm^4}{2p^4}+O(p^{-6})\right],
\label{frw6:vacuum-amplitude-difference}
\end{align}
which differs from Eq.~\eqref{frw6:inst-initial}.

The two normalized mode bases are related by
\begin{align}
u_p^{\rm inst}
&=\alpha_pu_p^{\rm A}+\beta_pu_p^{{\rm A}*},
\notag\\
\alpha_p
&=\frac12\left(\sqrt{\frac{W_{p0}}p}+\sqrt{\frac p{W_{p0}}}\right),
\qquad
\beta_p
=\frac12\left(\sqrt{\frac{W_{p0}}p}-\sqrt{\frac p{W_{p0}}}\right).
\label{frw6:vacuum-bogoliubov}
\end{align}
These coefficients are time independent and compare the two choices of vacuum. They satisfy $|\alpha_p|^2-|\beta_p|^2=1$, with
\begin{align}
\alpha_p=1+O(p^{-8}),\qquad
\beta_p=-\frac{qm^4}{2p^4}+O(p^{-6}).
\label{frw6:vacuum-beta-uv}
\end{align}

For Green functions and stress responses, the superscript $\mathrm{ins}$ denotes the instantaneous-vacuum result minus the adiabatic-vacuum result. We write $G_p^{R,K}[u]$ for the Green functions~\eqref{frw23:gr} and~\eqref{frw23:gk} evaluated with the mode function $u$. The retarded Green function is unchanged, while the Keldysh Green function changes by
\begin{align}
G_p^R[u^{\rm inst}](t,s)-G_p^R[u^{\rm A}](t,s)&=0,
\notag\\
G_p^K[u^{\rm inst}](t,s)-G_p^K[u^{\rm A}](t,s)
&=2|\beta_p|^2G_p^K[u^{\rm A}](t,s)
+4\operatorname{Re}\!\left[
\alpha_p\beta_p^*u_p^{\rm A}(t)u_p^{\rm A}(s)\right]
\notag\\
&=-\frac{qm^4}{p^5}\cos[p(t+s)]+O(p^{-6}).
\label{frw6:vacuum-Green-difference}
\end{align}
The additional phase depends on the sum of the two times. Inserting this term into the response kernel, the initial endpoint produces a logarithmic ultraviolet divergence. The integral over $\sigma=p'-p$ gives
\begin{align}
\int_{-k}^k d\sigma\,(k^2-\sigma^2)^2\cos(\sigma t)
=16k^5K(kt),
\end{align}
and the resulting divergent contribution is
\begin{align}
T_{ij}^{\rm dyn,ins,div}(t,\mathbf k)
=-\frac{\kappa qm^4}{4\pi^2\epsilon}
K(kt)h_{ij}(0,\mathbf k).
\label{frw6:vacuum-boundary-pole}
\end{align}
Here the logarithmic momentum integral is regulated as in Eq.~\eqref{frw5:radialpole}, and the observation time is fixed at $t>0$. Thus the change of state introduces a pole in the initial-value coefficient $F_1$.

The contact response also changes. Its ultraviolet contribution to the difference is
\begin{align}
T_{ij}^{\rm cont,ins}(t,\mathbf k)
=-\frac{\kappa qm^4}{4\pi^2}h_{ij}(t,\mathbf k)
\int_{p_0}^\infty\frac{dp}{p}\cos(2pt)
+\text{ultraviolet-finite terms},
\label{frw6:vacuum-contact-difference}
\end{align}
where $p_0>0$ is a lower limit in the ultraviolet region. At fixed $t>0$, this oscillatory integral has no ultraviolet pole, although it becomes logarithmically singular as $t\to0$. It therefore does not cancel Eq.~\eqref{frw6:vacuum-boundary-pole}, and the local bulk poles and their counterterms remain unchanged by this choice of state.

The initial-value divergence cannot be removed by the bulk counterterms. In the kinetic description, a contribution proportional to $K(kt)h_{ij}(0,\mathbf k)$ has a coefficient fixed by the initial condition for the stress tensor~\cite{Rebhan:1994zw,Ota:2025rll}. Although the full response has no exact Vlasov description, the high-momentum modes responsible for this pole free-stream at leading order, so the same homogeneous structure remains in the ultraviolet. We absorb the additional divergent constant into this coefficient: writing the homogeneous contribution as $C K(kt)h_{ij}(0,\mathbf k)$, we set
\begin{align}
C_{\rm bare}
=C_{\rm ren}+\frac{\kappa qm^4}{4\pi^2\epsilon}.
\label{frw6:vacuum-initial-renormalization}
\end{align}
The finite coefficient $C_{\rm ren}$ is specified by the initial condition. The two vacuum choices also give different finite responses. A calculation of the finite response therefore depends on the choice of initial vacuum.

\subsection{A Pauli--Villars regularization}
As a simple example, we apply Pauli--Villars regularization to the tadpole of the present model.
We introduce regulator fields with mass squared $U(t)+M_n^2$ and loop weights $Z_n^{-1}$~\cite{Armendariz-Picon:2019csc,Armendariz-Picon:2020tkc}. The physical field has $M_0=0$ and $Z_0=1$. Sums over $n=1,\ldots,N$ below include only the regulator fields. We impose
\begin{align}
\sum_{n=1}^N Z_n^{-1}=-1,\qquad
\sum_{n=1}^N Z_n^{-1}M_n^2=0,\qquad
\sum_{n=1}^N Z_n^{-1}M_n^4=0.
\label{frw5:pv-conditions}
\end{align}
Using a common momentum cutoff $\Lambda$, the regulated energy density and pressure are
\begin{align}
\rho_{\rm PV}
&=\frac1{4\pi^2}\sum_{n=0}^N Z_n^{-1}
\int_0^\Lambda dp\,p^2
\left[|\dot u_{n,p}|^2+(p^2+U+M_n^2)|u_{n,p}|^2\right],
\notag\\
P_{\rm PV}
&=\frac1{4\pi^2}\sum_{n=0}^N Z_n^{-1}
\int_0^\Lambda dp\,p^2
\left[|\dot u_{n,p}|^2-
\left(\frac{p^2}{3}+U+M_n^2\right)|u_{n,p}|^2\right].
\label{frw5:pv-mode-integrals}
\end{align}
Here $u_{n,p}$ is the mode function for mass squared $U+M_n^2$, and a dot denotes a derivative with respect to $t$. We use the adiabatic expansion with the regulator masses retained in the zeroth-order frequency, $\omega_{n,p}^2=p^2+U+M_n^2$. The conditions in Eq.~\eqref{frw5:pv-conditions} cancel the cutoff dependence, allowing $\Lambda\to\infty$ at fixed regulator masses. We then take the regulator masses to infinity at fixed ratios.

To compare with the poles above, define
\begin{align}
\mathcal I_{\rm PV}
=\sum_{n=1}^N Z_n^{-1}\log\!\left(\frac{M_n}{\mu}\right).
\label{frw5:pv-log}
\end{align}
The common logarithmic contributions are
\begin{align}
\rho_{\rm PV,log}
&=\frac{m^4q^2\sin^4(mt)}{2\pi^2}\mathcal I_{\rm PV},
\notag\\
P_{\rm PV,log}
&=\left[\frac{m^4q\cos(2mt)}{6\pi^2}
-\frac{m^4q^2\sin^4(mt)}{2\pi^2}\right]\mathcal I_{\rm PV}.
\label{frw5:pv-log-tadpole}
\end{align}
Terms proportional to $\sum_{n=1}^N Z_n^{-1}M_n^2\log(M_n/\mu)$ and $\sum_{n=1}^N Z_n^{-1}M_n^4\log(M_n/\mu)$ represent power divergences and are absorbed into local counterterms $c_2$ and $c_1$. Equation~\eqref{frw5:pv-log-tadpole} displays the remaining common logarithmic structure. The flat-background limits of Eqs.~\eqref{frw5:rhopole} and~\eqref{frw5:Ppole} agree with it under the replacement
\begin{align}
\frac1\epsilon\ \longleftrightarrow\ -\mathcal I_{\rm PV}.
\label{frw5:pv-pole-matching}
\end{align}
This comparison concerns the logarithmic terms; the finite parts depend on the regularization and renormalization conditions.

Similarly, for the $r=2$ response, we apply the same weighted sum over
species to the Green's function product in Eq.~(2.19).
The four-derivative coefficient is independent of $U(t)$,
so we set $U=0$ in this calculation and retain the regulator
masses $M_n$. The local logarithmic term is
\begin{align}
T_{ij,\mathrm{PV,log}}^{(2)}
=\frac{\kappa I_{\mathrm{PV}}}{960\pi^2}\Box^2h_{ij}.
\label{frw5:r=2-pv-matching}
\end{align}
This agrees with Eq.~(6.11) under the identification in Eq.~(6.34).

\section{Conclusions}\label{frw6:concl}

In this paper we have investigated the renormalization of the stress tensor in the real-time linear response of gravitational waves in a scalar field with time-dependent mass in an FLRW background, where the scalar field has a general time-dependent effective mass squared $U(\tau)$. Our main purpose has been to clarify the ultraviolet structure of the one-loop response and to identify the local counterterms required for its renormalization. To this end, we formulated the problem in the Schwinger-Keldysh formalism, adopted fourth-order adiabatic initial conditions, and developed an adiabatic regularization scheme adapted to unequal-time correlation functions in a time-dependent setting.

In the present problem, the adiabatic expansion cannot be identified naively with the ultraviolet expansion. Although the large-momentum regime satisfies the adiabatic conditions of the time-dependent background, the ultraviolet behavior of unequal-time Green functions is controlled not only by the large-$p$ asymptotics of the mode functions but also by the short-distance structure in the time separation. For this reason, the coincidence limit and the large-momentum limit must be taken in a correlated way. By introducing an auxiliary bookkeeping parameter for the simultaneous expansion in inverse momentum and time separation, we obtained a systematic asymptotic expansion of the retarded and Keldysh Green functions that enter the linear-response kernel.

This construction makes it possible to isolate the divergent part of the one-loop response systematically. At leading ultraviolet order, we recover the familiar four-derivative divergent structure $\Box^2h_{ij}$. This term is renormalized by the Weyl-squared counterterm, in agreement with the standard expectation. At higher orders, the explicit time dependence of the background and the FLRW curvature give rise to additional local structures proportional to $\Box h_{ij}$, $\partial_\tau h_{ij}$, and $h_{ij}$ itself. We have shown that these divergences can be absorbed into the Ricci-scalar and cosmological-constant counterterms with time-dependent coefficients, together with the Ricci-scalar-squared counterterm, as summarized in table~\ref{frw5:uv-summary}. Thus, the ultraviolet divergences of a real-time gravitational-wave response can still be organized systematically even in the absence of time-translation invariance.

We have also examined the relation between the renormalization of the linear response and that of the tadpole stress tensor. Once the contact response is included, the counterterms required by the complete linear response agree with those required by the tadpole in the FLRW background. The coefficients $c_1(U)$, $c_2(U)$, and $c_5$ determined independently from the tadpole coincide with those obtained from the linear response. Since the Weyl tensor vanishes in the FLRW background, $c_3$ is determined by the four-derivative structure of the linear response and cannot be checked independently from the background tadpole.

From this point of view, the present analysis shows that the local ultraviolet divergences induced by a time-dependent effective mass in an FLRW background can be treated consistently with the same counterterms in the linear response and the tadpole. The resulting counterterms agree with the local DeWitt--Schwinger expansion, and the covariant Pauli--Villars cross-check yields the same common logarithmic coefficients.

The initial state is a separate issue. In the flat-background Mathieu model, the instantaneous vacuum at $t_0=0$ differs from the adiabatic vacuum by a Bogoliubov coefficient $\beta_p=O(p^{-4})$. At fixed $t>0$ the local poles and their counterterms are unchanged, but the response acquires a divergence proportional to $K(kt)h_{ij}(0,\mathbf k)$. It is not removed by the bulk counterterms and is absorbed into the coefficient of the initial homogeneous contribution, and the finite response depends on the choice of initial vacuum.

The present work is a step toward a renormalized real-time effective theory of gravitational waves in nonequilibrium time-dependent backgrounds. The main result of this paper is not merely the identification of the divergent terms themselves, but the demonstration that their structure can be extracted systematically from unequal-time real-time correlators by combining adiabatic and ultraviolet expansions in an appropriate way. At the same time, the physically more interesting part of the response is the finite nonlocal kernel, which encodes the actual modification of gravitational-wave propagation by the time-dependent background. Determining this finite contribution beyond the adiabatic approximation, and clarifying its relation to particle production, secular growth, and nonadiabatic effects, remain problems for future work.

\begin{table}[!htbp]
\centering
\small
\setlength{\tabcolsep}{5pt}
\renewcommand{\arraystretch}{1.2}
\setlength{\extrarowheight}{2pt}
\begin{tabular}{@{}ccccc@{}}
\hline
\noalign{\vskip 4pt}
UV order in $r$ & Linear response & Counterterm & Interpretation & Ref. \\[4pt]
\hline
\noalign{\vskip 6pt}
$2$
& $\Box ^2h_{ij}$
& $c_3C_{\mu\nu\rho\sigma}C^{\mu\nu\rho\sigma}$
& \shortstack[l]{four-derivative\\curvature term}
& Eq.~\eqref{frw4:c3} \\
\noalign{\vskip 8pt}
$4,5$
& $\Box h_{ij}$, $h'_{ij}$
& $c_2(U)R$, $c_5R^2$
& \shortstack[l]{time-dependent\\gravitational coupling\\and curvature-squared term}
& Eq.~\eqref{frw4:c2c5} \\
\noalign{\vskip 8pt}
\shortstack{$6$ and residual\\terms from $4,5$}
& $h_{ij}$
& $c_1(U)$
& \shortstack[l]{time-dependent\\cosmological constant term}
& Eq.~\eqref{frw4:c1} \\
\noalign{\vskip 6pt}
\hline
\end{tabular}
\captionsetup{skip=9pt}
\caption{Summary of the divergent local structures in the linear response and the corresponding counterterms.
The $r=5$ term is induced by the time dependence of $c_2$ and the curvature coefficient.
The $r=6$ row denotes the residual local mass-like term after including the contact contribution and subtracting the divergences absorbed into $c_3$, $c_2$, and $c_5$.
In a general time-dependent background, $c_1$ and $c_2$ are functions of $U$, while $c_3$ and $c_5$ are constants.}
\label{frw5:uv-summary}
\end{table}

\acknowledgments
This work was supported in part by the National Natural Science Foundation of China under Grant No. 12347101 and 12403001, and New Chongqing YC Project CSTB2024YCJH-KYXM0083.
AO is grateful to Yingying Lan for her generous support.
During the preparation of this work, the authors used the AI tools GPT-6.1 Sol (OpenAI) and Claude Opus 5.5 (Anthropic) to assist with cross-checks of calculations, code for numerical checks, and editing of the text. The authors reviewed and verified all results and text obtained with these tools and take full responsibility for the content of this paper.

\clearpage
\bibliographystyle{unsrturl}
\bibliography{biblio}
\end{document}